\documentclass[prb,twocolumn,superscriptaddress,showpacs,amsmath,amssymb]{revtex4}
\usepackage{amsfonts}
\usepackage{bbm}
\usepackage{graphicx}% Include figure files
\usepackage{dcolumn}% Align table columns on decimal point
\usepackage{bm}% bold math
\usepackage{subfigure}
\usepackage{color}

\begin{document}
\title{Controllable Josephson diode effect, $0$-$\pi$ transition and switch effect in the superconductor/two-dimensional Weyl nodal line semimetal/superconductor junctions}

\author{Wen-Ting Liu}
\affiliation{School of Science, Qingdao University of Technology, Qingdao, Shandong 266520, China}

\author{Shu-Chang Zhao}
\affiliation{School of Science, Qingdao University of Technology, Qingdao, Shandong 266520, China}

\author{Qiang Cheng}
\email[]{chengqiang07@mails.ucas.ac.cn}
\affiliation{School of Science, Qingdao University of Technology, Qingdao, Shandong 266520, China}
\affiliation{International Center for Quantum Materials, School of Physics, Peking University, Beijing 100871, China}
	
\author{Qing-Feng Sun}
\email[]{sunqf@pku.edu.cn}
\affiliation{International Center for Quantum Materials, School of Physics, Peking University, Beijing 100871, China}
\affiliation{Hefei National Laboratory, Hefei 230088, China}

\begin{abstract}
We study the Josephson effects in the superconductor/two-dimensional Weyl nodal line semimetal/superconductor junctions using the Green's function method. When the Rashba spin-orbit coupling and an external magnetic field coexist in the semimetal, the symmetries protecting the reciprocity of supercurrent can be broken and the Josephson diode effect with the nonreciprocity of supercurrent can be realized. A high efficiency exceeding $40\%$ can be achieved with the experimentally accessible values of the magnetic field and the Rashba spin-orbit coupling. The diode efficiency can be easily controlled by the direction and magnitude of the field and the strength of the spin-orbit coupling. When the spin-orbit coupling is absent or the external field is absent, the Josephson diode effect vanishes but the current-phase difference relations still show strong dependence on the field or the coupling. The tunable $0$-$\pi$ transition and the switch effect of supercurrent in the junctions can be formed if the direction of the field is rotated or the magnitude of the field and the strength of the coupling are changed. The obtained Josephson diode effect, the $0$-$\pi$ transition and the switch effect of supercurrent are helpful in the design of the quantum devices based on nodal line semimetals.
\end{abstract}
\maketitle
	
\section{\label{sec1}Introduction}
The Josephson effect is a basic physical phenomenon, which can host supercurrent without resistance in a Josephson structure consisting of the non-superconducting material sandwiched between two superconductors (SCs).
The critical current is the maximum supercurrent that a Josephson structure can sustain. If one defines the value of the critical current for the positive direction as $I_{C+}$ and that for the negative direction as $I_{C-}$, the relation $I_{C+}=I_{C-}$ is usually satisfied due to the protection of symmetries, e.g., the time-reversal symmetry or the inversion symmetry. However, the relation can be violated when symmetries obeyed by the Josephson structure are broken, which will lead to the Josephson diode effect (JDE) with $I_{C+}\ne I_{C-}$. Recently, JDEs have been experimentally realized in various Josephson structures\cite{Wu,Pal,Baumgartner,Banerjee,Lin,Trahms,Matsuo}, such as the van der Waals heterostructure\cite{Wu}, the type-II Dirac semimetal junctions\cite{Pal}, the InAs quantum wells junctions\cite{Baumgartner,Banerjee} and the small-twist-angle trilayer graphene junctions\cite{Lin}. Among them, experiments in Refs.[\onlinecite{Wu,Pal,Lin}] demonstrate the stable half-wave rectification of JDEs, which paves the way for the design of dissipationless superconducting quantum devices.

Simultaneously, many theoretical schemes for JDE have also been proposed\cite{Halterman,Davydova,Fominov,YZhang,Souto,Tanaka,Kokkeler,Cheng1,YFSun1,Cheng2,YFSun2,Lu,Maoadd,Costa,Cayao,Maiani,Vakili,Seleznev,
Zazunov,Legg,Ding,Patil,Ilic,Scharf,JXHu}. For example, the influences of the Rashba spin-orbit coupling (RSOC) and the magnetic exchange field on the current-phase difference relations (CPRs) in the two-dimensional electron gas Josephson junctions are clarified by the authors in Ref.[\onlinecite{Costa}]. The realization of JDE in the junctions and its tunability in terms of RSOC and the exchange field are discussed. Furthermore, the current-reversing $0$-$\pi$-like transitions are found for a large enough exchange field. In Ref.[\onlinecite{Cayao}], the authors study the Josephson effect in nanowire junctions with RSOC under external magnetic fields. It is found that JDE emerges when the field in the middle region is along the RSOC axis, which can be enhanced by the presence of Majorana bound states. In the Josephson junctions based on semiconductor-superconductor-ferromagnetic insulator heterostructures\cite{Maiani}, the control of the supercurrent harmonics and the $0$-$\pi$ transition with the supercurrent reversal are investigated. The JDE can be achieved in the junctions when the condition of noncollinear magnetization or noncollinear RSOCs is met. For the two-dimensional diffusive Josephson junctions with RSOC\cite{Ilic}, the increase of the JDE efficiency and its sign change close to the $0$-$\pi$ transition can occur at specific magnitudes of the magnetic field.

On the other hand, the heterojunctions consisting of the nodal line semimetal (NLSM) and SC exhibit peculiar transport properties. For the three-dimensional situation with the torus-shaped Fermi surface, double normal reflections, double Andreev reflections, double electron transmissions, double crossed Andreev reflections or the anomalous crossed Andreev reflection with one lateral velocity component have recently been demonstrated\cite{Cheng3,YXWang,XWang}. For the two-dimensional case with the line node and the ring-shaped Fermi surface, the pure specular Andreev reflection and the resulting unique transport properties distinct from those for the retro-Andreev reflection are clarified\cite{Cheng4}. As a result, it is believed that the NLSM Josephson junctions can host peculiar Josephson effects. Especially, the gigantic magnetochiral anisotropy with the nonreciprocal charge transport has been recently observed in the inversion breaking three-dimensional NLSM ZrTe$_5$ under a magnetic field along the crystallographic $b$ axis\cite{YWang}. Therefore, JDE can be expected in the SC/NLSM/SC junctions when the inversion and time-reversal symmetries are broken simultaneously.
	
In this paper, we study the Josephson effects in the SC/NLSM/SC junctions and propose a theoretical scheme for JDE realized in the junctions.  For the convenience of calculation, the involved NLSM is chosen as the two-dimensional Weyl NLSM used in Ref.[\onlinecite{Cheng4}]. Several candidate materials are proposed for this type of NLSM based on theoretical calculations\cite{Feng,Nie,Jin,Niu}. When an external field is applied to NLSM, the CPRs in our junctions strongly depend on the direction and magnitude of the field. The $0$-$\pi$ transition and the supercurrent switch effect can be achieved by rotating the orientation of the field or increasing the magnitude of the field. When RSOC is introduced in NLSM, the CPRs in the junctions also strongly depend on the strength of RSOC and the supercurrent switch effect can also be realized when the strength of RSOC is changed. When RSOC and the external field coexist in NLSM, JDE with the supercurrent nonreciprocity will emerge. Its efficiency can exceed $40\%$ and can be controlled by adjusting the direction of the field and the magnitudes of the field and RSOC. The JDE and the supercurrent switch effect found in our junctions possess the potential applications in the design of dissipationless devices and the $0$-$\pi$ transition can be used for the circuit elements of the superconducting computer.

The organization of this paper is as follows. In Sec. $\text{\uppercase\expandafter{\romannumeral2}}$, the model of the proposed junctions and the Green's function formalism are presented. In Sec. $\text{\uppercase\expandafter{\romannumeral3}}$, the numerical results are discussed. Sec. $\text{\uppercase\expandafter{\romannumeral4}}$ concludes this paper. The details about the electronic structure of NLSM, the discretization of Hamiltonians and the derivation of the Green's functions are given in the Appendix.

\section{\label{sec2}Model and Formulation}

\begin{figure}[!htb]
\centerline{\includegraphics[width=1\columnwidth]{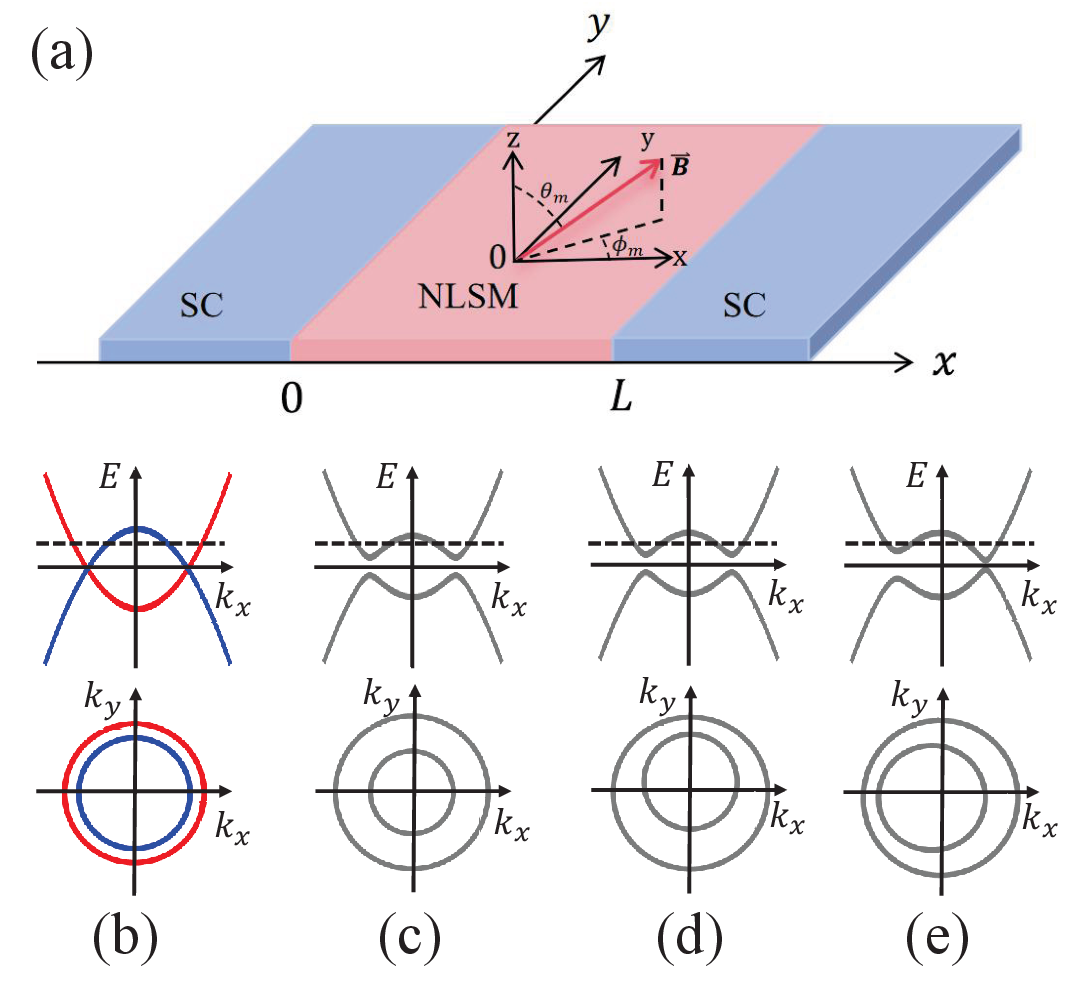}}
\caption{(a) The schematic illustration of the SC/NLSM/SC junctions. The external magnetic field is denoted by the red arrow, whose direction is described by the polar angle $\theta_m$ and the azimuthal angle $\phi_m$. The energy bands and Fermi surfaces of electrons in NLSM are schematically shown for (b) $\lambda=0$ and $B_{\parallel}=0$, (c) $\lambda=0$, $B_{\parallel}\ne0$ or $\lambda\ne0$, $B_{\parallel}=0$, (d) $\lambda\ne0$, $B_y=0$ and $B_x\ne0$, and (e) $\lambda\ne0$, $B_x=0$ and $B_y\ne0$. The black dotted lines denote the positions of the Fermi surfaces. The red (blue) band and red (blue) Fermi surface in (b) are for the spin-up (spin-down) electrons. Here, $B_{\parallel}=(B_x,B_y)$ is the in-plane component of the external field ${\bf{B}}$.}\label{fig1}
\end{figure}

The SC/NLSM/SC junctions considered by us are schematically shown in Fig.\ref{fig1}, which are placed in the $xy$ plane. The current flows along the direction parallel to the $x$ axis. The interfaces of the junctions are located at $x=0$ and $x=L$ and $L$ denotes the length of the middle NLSM. The left lead and the right lead belong to the conventional SC with the isotropic $s$-wave pairing. The Bogoliubov-de Gennes (BdG) Hamiltonian for the left (right) SC can be expressed as
\begin{eqnarray}
H_{L(R)}=\sum_{\bf{k}}\Psi_{L(R)}^{\dagger}({\bf{k}})\check{H}_{L(R)}({\bf{k}})\Psi_{L(R)}({\bf{k}}),\label{1}
\end{eqnarray}
with $\Psi_{L(R)}({\bf{k}})=(c_{L(R){\bf{k}}\uparrow},c_{L(R){\bf{k}}\downarrow},c_{L(R){-\bf{k}}\uparrow}^{\dagger},c_{L(R){-\bf{k}}\downarrow}^{\dagger})^{T}$ in the particle-hole$\otimes$spin space. Here, ${\bf{k}}=(k_x,k_y)$ is the two-dimensional wavevector and the single-particle Hamiltonian is expressed as
\begin{eqnarray}
\begin{split}
\check{H}_{L(R)}({\bf{k}})&=\epsilon_{L(R)}\tau_z\otimes\sigma_0\\
&-\Delta_{0}[\cos{\phi_{l(r)}}\tau_{y}+\sin{\phi_{l(r)}}\tau_{x}]\otimes\sigma_{y},\label{2}
\end{split}
\end{eqnarray}
with kinetic energy $\epsilon_{L(R)}=t_0 (k_x^2+k_y^2)-\mu_{L(R)}$ and $t_0=\frac{\hbar^2}{2m}$. We will take $\hbar=1$ in the following. The symbols $\mu_{L(R)}$ and $\Delta_0$ denote the chemical potential in the left (right) SC and the gap magnitude in SCs. The superconducting phases in the left SC and the right SC are represented by $\phi_l$ and $\phi_r$, respectively. We have used $\tau_x,\tau_y,\tau_z$ and $\sigma_x,\sigma_y,\sigma_z$ to denote the Pauli matrices in the particle$\otimes$hole space and the spin space while $\tau_0$ and $\sigma_0$ are the corresponding identity matrices.
	
The BdG Hamiltonian for the middle NLSM can be expressed as\cite{Cheng4,Jin}
\begin{eqnarray}
H_{M}=\sum_{\bf{k}}\Psi_{M}^{\dagger}({\bf{k}})\check{H}_{M}({\bf{k}})\Psi_{M}({\bf{k}}),\label{5}
\end{eqnarray}
with $\Psi_{M}({\bf{k}})=(c_{M{\bf{k}}\uparrow},c_{M{\bf{k}}\downarrow},c_{M{-\bf{k}}\uparrow}^{\dagger},c_{M{-\bf{k}}\downarrow}^{\dagger})^{T}$ and the single-particle Hamiltonian is given by
\begin{eqnarray}
\begin{split}
\check{H}_{M}({\bf{k}})&=\epsilon_{M}\tau_z\otimes\sigma_z\\
&+B_{x}\tau_{z}\otimes\sigma_{x}+B_{y}\tau_{0}\otimes\sigma_{y}+B_{z}\tau_{z}\otimes\sigma_{z}\\
&+\lambda(k_{y}\tau_{0}\otimes\sigma_{x}-k_{x}\tau_{z}\otimes\sigma_{y}),\label{NLSMH}
\end{split}
\end{eqnarray}
with $\epsilon_{M}=t_0 (k_x^2+k_y^2)-\mu_{M}$. Here, $\mu_{M}$ is the quantity characterizing the size of the nodal line in NLSM. The external field is expressed as ${\bf{B}}=(B_x,B_y,B_z)=B{\bf{n}}$ with $B$ being its magnitude. The direction of the magnetic field is characterized by a unit vector ${\bf{n}}=(\sin{\theta_m}\cos{\phi_m},\sin{\theta_m}\sin{\phi_m},\cos{\theta_m})$ in the three-dimensional coordinate system with $\theta_m$ being its polar angle and $\phi_m$ being its azimuthal angle (see Fig.\ref{fig1}). The last term in Eq.(\ref{NLSMH}) is RSOC introduced in NLSM and $\lambda$ is its strength.
The detailed derivation process of Eq.(\ref{NLSMH}) is provided in Appendix A.
The energy bands and Fermi surfaces of electrons in NLSM under RSOC and the external magnetic field are schematically shown in Figs.{\ref{fig1}}(b)-(e), which are derived from the Hamiltonian of NLSM in the spin space (see Appendix A).
Their characteristic properties are also described in Appendix A.
	
In order to calculate the Josephson current in the junctions, we discretize the Hamiltonians $H_{L(R)}$ and $H_{M}$ on a two-dimensional square lattice with the lattice constant $a$ (see Appendix B). The discrete Hamiltonian for the left (right) SC is given by
\begin{eqnarray}
%\begin{aligned}
H_{L(R)}&=&\sum_{{\bf{i}}}\left[\Psi_{{{L(R)\bf{i}}}}^{\dagger}\check{H}_{L(R)0}\Psi_{{{L(R)}\bf{i}}}
\right.\nonumber\\
&&
\left.+\Psi_{{{L(R)}\bf{i}}}^{\dagger}\check{H}_{L(R)x}\Psi_{{{L(R)}{\bf{i}}+\delta x}}
\right.\nonumber\\
&&
\left.+\Psi_{{{L(R)}\bf{i}}}^{\dagger}\check{H}_{L(R)y}\Psi_{{{L(R)}{\bf{i}}+\delta y}}+H.C.\right],\label{10}
%\end{aligned}
\end{eqnarray}
with $\Psi_{{L(R)}{\bf{i}}}=(\Psi_{{L(R)\bf{i}}\uparrow},\Psi_{{L(R)\bf{i}}\downarrow}, \Psi_{{L(R)\bf{i}}\uparrow}^{\dagger}, \Psi_{{ L(R)\bf{i}}\downarrow}^{\dagger})^{T}$.
Here the subscript ${\bf{i}}=(i_x,i_y)$ refers to the coordinate position of the ${\bf{i}}$th lattice point in a two-dimensional square lattice, the subscript ${\bf{i}}+\delta x$ represents the nearest-neighbor lattice point along the x direction and the subscript ${\bf{i}}+\delta y$ indicates the nearest-neighbor lattice point along the y direction.
The discrete Hamiltonian for the middle NLSM can be given by
\begin{eqnarray}
\begin{split}
H_{M}=\sum_{\bf{i}}[\Psi_{{{M}\bf{i}}}^{\dagger}\check{H}_{M0}\Psi_{{{M}\bf{i}}}+\Psi_{{{M}\bf{i}}}^{\dagger}\check{H}_{Mx}\Psi_{{{M}{\bf{i}}+\delta x}}\\
+\Psi_{{{M}\bf{i}}}^{\dagger}\check{H}_{My}\Psi_{{{M}{\bf{i}}+\delta y}}+H.C.].\label{11}
\end{split}
\end{eqnarray}
with $\Psi_{{{M}\bf{i}}}=(\Psi_{M{\bf{i}}\uparrow},\Psi_{M{\bf{i}}\downarrow},\Psi_{M{\bf{i}}\uparrow}^{\dagger},\Psi_{M{\bf{i}}\downarrow}^{\dagger})^{T}$. If the width $W$ and the length $L$ of the junctions are given by $W=(w-1)a$ and $L=(l-1)a$, the range of $i_y$ will be constrained to $1 \le i_y \le w$ and the range of $i_x$ will be $i_x \le 0$, $1 \le i_x \le l$,
and $l+1 \le i_x$ for the left SC, the middle NLSM, and the right SC, respectively.
The derivation of the matrices $\check{H}_{L(R)0}$, $\check{H}_{L(R)x}$, $\check{H}_{L(R)y}$,$\check{H}_{M0}$, $\check{H}_{Mx}$ and $\check{H}_{My}$ in Eq.(\ref{10}) and Eq.(\ref{11}) will be discussed in Appendix B.
	
The Hamiltonian describing the hopping between the left SC and NLSM and between the right SC and NLSM can be expressed as
\begin{eqnarray}
H_{T} &= &\sum_{1\le i_y\le w}\left[\Psi_{L(0,i_y)}^{\dagger}\check{T}_{L}\Psi_{M(1,i_y)} \right.\nonumber\\
		&& \left.+\Psi_{R(l+1,i_y)}^{\dagger}\check{T}_{R}\Psi_{M(l,i_y)}+H.C.\right].\label{12}
\end{eqnarray}
The subscript $(0,i_y)$ denotes the coordinates of lattice points in the rightmost column of the left SC, $(1,i_y)$ denotes the coordinates of lattice points in the leftmost column of middle NLSM, $(l,i_y)$ denotes the coordinates for points in the rightmost column of middle NLSM and $(l+1,i_y)$ denotes the coordinates for points in the leftmost column of the right SC.
The hopping matrices $\check{T}_{L}$ and $\check{T}_{R}$ are given by $\check{T}_{L(R)}=\text{diag}(t,t,-t,-t)$.
	
According to the definition of the particle number operator within the framework of quantum many-body theory, we define the particle number operator for the left superconducting region as 	
\begin{eqnarray}
N_L=\sum_{i_x\le0,i_y,\sigma}\Psi_{L\bf{i}\sigma}^{\dagger}\Psi_{L\bf{i}\sigma}.\label{15}
\end{eqnarray}
Then, the Josephson current can be written as\cite{Cheng2,Cheng5,Cheng6}
\begin{equation}
\begin{aligned}
I&=e\langle \frac{dN_L}{dt}\rangle\\
&=-\frac{e}{2\pi}\int dE \text{Tr}[\Gamma_z\check{T}_{L}G_{ML}^{<}(E)+H.C.].\label{16}
\end{aligned}
\end{equation}
with $\Gamma_{z} = \sigma_z \otimes 1_{2\times2}$ and $G_{ML}^{<}(E)$ being the lesser Green's function. The derivation of the lesser Green's functions can be found in Appendix C.
	
\section{\label{sec3}Numerical results and discussions}
In our calculations, we take $t_0=1$ and the chemical potential in SCs is taken as $\mu=2$. The magnitude of the hopping between the left SC or the right SC and NLSM is taken as $t=1$. The parameter characterizing the size of the nodal line in NLSM is taken as $\mu_{M}=2$. The lattice parameters are taken as $a=1$, $N_x=10$ and $N_y=10$. The phase difference is defined as $\phi=\phi_l-\phi_r$. The unit of the Josephson current is taken as $e\Delta_0/2\pi$ with the gap magnitude $\Delta_0$ being $0.01$. Next, we will discuss the numerical results in three subsections. The first subsection is for the junctions with the external field and without RSOC, i.e., $B\ne0$ and $\lambda=0$. The second subsection is for the junctions with RSOC and without the external field, i.e., $\lambda\ne0$ and $B=0$. The third subsection is for the junctions with both the external field and RSOC, i.e., $B\ne0$ and $\lambda\ne0$.

\subsection{$B\ne0$ and $\lambda=0$}
\begin{figure}[!htb]
\centerline{\includegraphics[width=0.9\columnwidth]{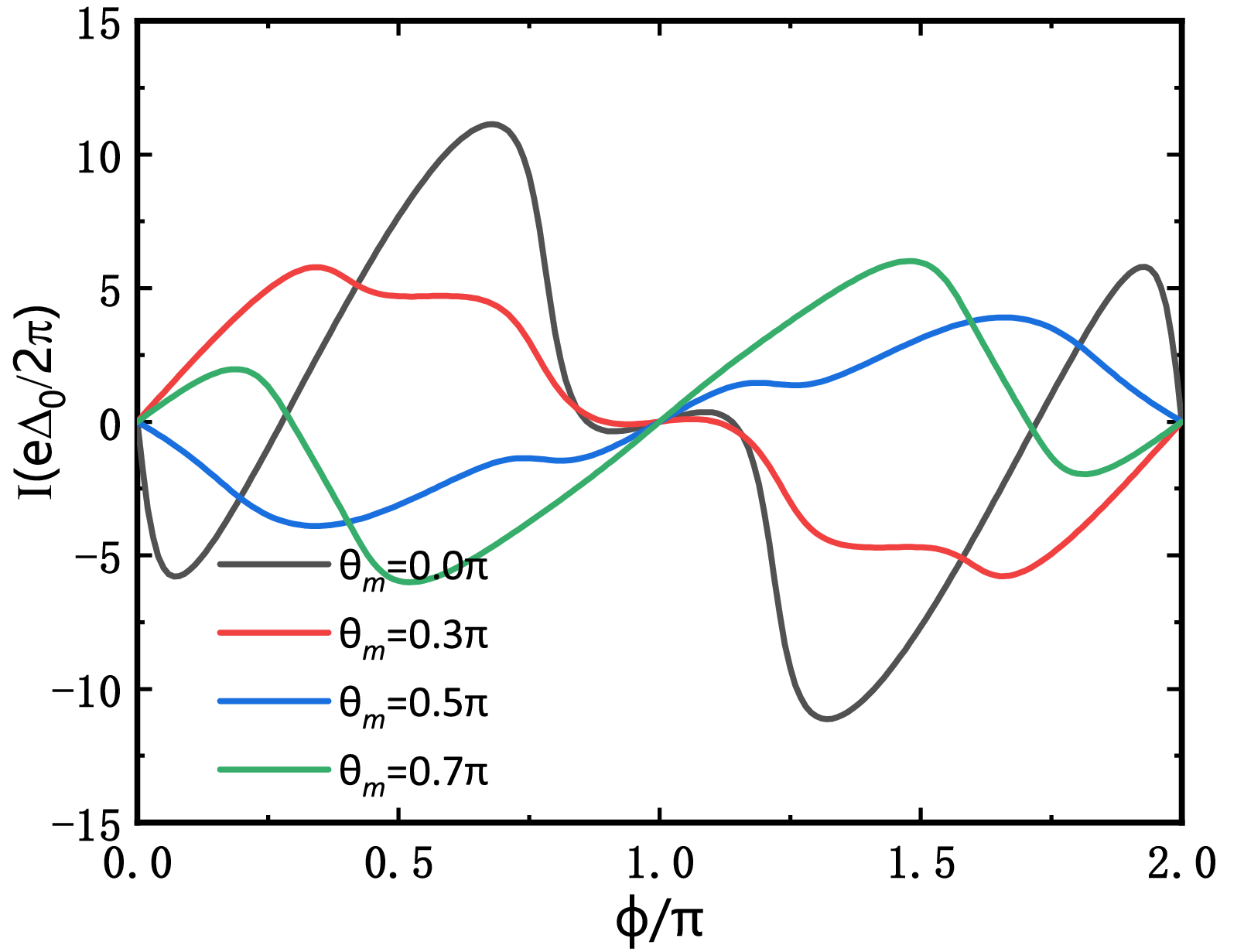}}
\caption{CPRs for different polar angles of the external field. The azimuthal angle of the field is taken as $\phi_{m}=0.5\pi$ and the magnitude of the field is taken as $B=0.3$.}\label{fig2}
\end{figure}

In this situation, the junctions preserve the spin-rotation symmetry about the $z$ axis and the Josephson current is independent of the azimuthal angle $\phi_{m}$ of the field. However, CPRs are strongly dependent on the polar angle $\theta_m$ of the field as shown in Fig.\ref{fig2}. From Fig.\ref{fig2}, it is found that the critical current of CPRs gradually decreases from its positive value to its negative value when the polar angle of the field is changed from $\theta_m=0$ to $\theta_m=0.7\pi$. The positive critical current means that the minimum value of the free energy for the junctions is reached at $\phi=0$ while the negative critical current means that the minimum value is reached at $\phi=\pi$. In other words, the $0$-$\pi$ transition will occur as the polar angle is increased from $0$ to $0.7\pi$. Therefore, our junctions can realize the easily controllable $0$-$\pi$ transition by rotating the external field. The $0$-$\pi$ transition is first theoretically proposed by Buzdin et al.\cite{Buzdin} and has been observed in experiment\cite{Ryazanov,Kontos}, which can be served as the necessary component of the energy-efficient programmable logic device based on SCs\cite{Gingrich}.

\begin{figure}[!htb]
\centerline{\includegraphics[width=1\columnwidth]{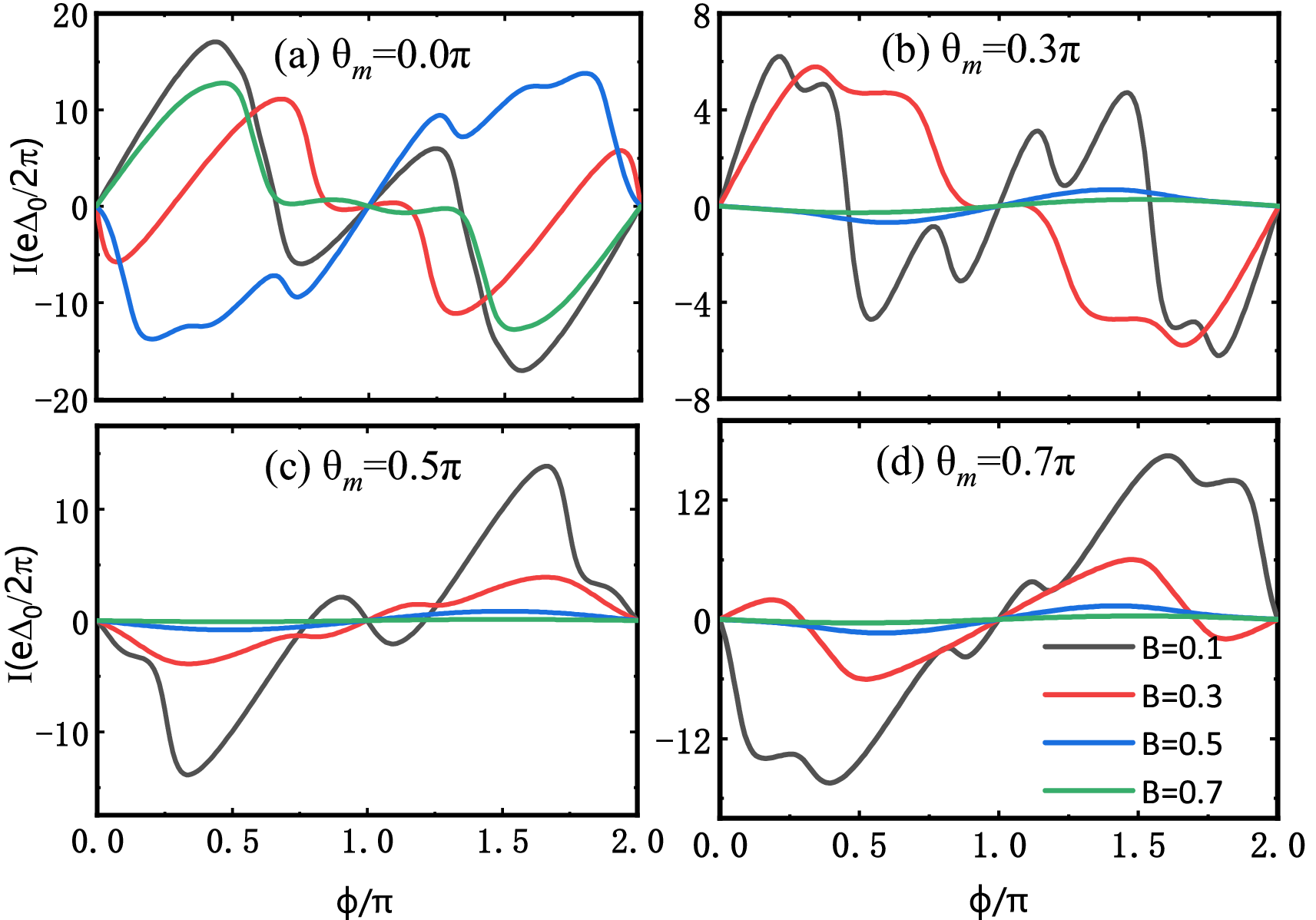}}
\caption{CPRs for different magnitudes of the external field with (a) $\theta_m$=0, (b) $\theta_m=0.3\pi$, (c) $\theta_m=0.5\pi$, and (d) $\theta_m=0.7\pi$. The azimuthal angle is taken as $\phi_m=0.5\pi$.}\label{fig3}
\end{figure}

In Fig.\ref{fig3}, we present CPRs for different magnitudes of the external field. For the polar angles $\theta_m=0$ and $0.3\pi$,
the obvious $0$-$\pi$ transition can also be observed when the magnitude of the field is increased from $B=0.1$ to $B=0.7$, which provides another method to tune the $0$-$\pi$ transition. For $\theta_m=0.5\pi$, i.e., the external field lying on the $xy$ plane, the increase of the magnitude of the field does not produce the $0$-$\pi$ transition. Furthermore, the critical current will vary from the finite value for $B=0.1,0.3,0.5$ to a value close to zero for $B=0.7$. In other words, one can switch the Josephson current through adjusting the magnitude of the field. Actually, the switch of the Josephson current can also be realized by rotating the field with $B=0.7$ from $\theta_m=0$ to $\theta_m=0.5\pi$ as shown in Figs.\ref{fig3}(a)-(c). If we define the switch efficiency as
\begin{eqnarray}
\kappa=\frac{\vert I(\theta_m=0)\vert-\vert I(\theta_m=\pi/2)\vert}{\vert I(\theta_m=0)\vert+\vert I(\theta_m=\pi/2)\vert},
\end{eqnarray}
its maximum value for $B=0.7$ can exceed $98\%$. The switch effect here originates from the peculiar electronic structure of the two-dimensional Weyl NLSM. The presence of the in-plane component of the field, i.e., $B_x$ or $B_y$, will open a gap in the energy bands of NLSM. When the in-plane component of the field is large enough, the Josephson current will vanish. This effect in our junctions can not be expected in the SC/normal metal/SC junctions under an external field or the SC/ferromagnet/SC junctions with an exchange field. CPRs in these junctions remain unchanged for the rotation of the external field or the exchange field.

For $B\ne0$ and $\lambda=0$ here, JDE does not exist in our junctions, which can be observed from CPRs in Figs. \ref{fig2} and \ref{fig3}.
This can also be inferred from the energy bands and Fermi surfaces in Figs.{\ref{fig1}}(b) and (c), which are isotropic in the $(k_x,k_y)$ space.
Generally speaking, the Josephson current can be decomposed into the Fourier series, i.e., $I(\phi)=\sum_{n\ge1}[a_n\sin{n\phi}+b_n\cos{n\phi}]$ with $n$ being an integer number. From Figs. \ref{fig2} and \ref{fig3}, one can find that CPRs only include the $\sin{n\phi}$ term since one has $I=0$ at $\phi=0,\pi$ and $2\pi$. However, the formation of JDE is usually accompanied by the presence of the $\cos{\phi}$ term\cite{Ilic,JXHu,Vakili,Cheng2,Baumgartner}. The absence of JDE in this situation can be understood through the symmetry analysis. We introduce the inversion-rotation joint operation denoted by $\mathcal{U}_i=\mathcal{M}_{xz}\mathcal{M}_{yz}\mathcal{R}_{z}(\pi)$ with $\mathcal{M}_{xz}$ and $\mathcal{M}_{yz}$ being the mirror reflection operations about the $xz$ plane and the $yz$ plane and $\mathcal{R}_{z}(\pi)$ being the $\pi$-angle spin-rotation about the $z$ axis\cite{Cheng2}. The action of $U_{i}$ on the Nambu spinor $(c_{\bf{k}\uparrow},c_{\bf{k}\downarrow},c^{+}_{-\bf{k}\uparrow},c^{+}_{-\bf{k}\downarrow})$ can be expressed as $U_{i}c_{{\bf{k}\sigma}}U_{i}^{-1}=c_{-{\bf{k}}\sigma}$ and $U_{i}c^{+}_{{\bf{k}\sigma}}U_{i}^{-1}=c^{+}_{-{\bf{k}}\sigma}$. The inversion-rotation joint operation can transform the Hamiltonian for the left (right) SC in the following manner
\begin{eqnarray}
\mathcal{U}_{i}H_{L(R)}(\phi_{L(R)})\mathcal{U}_{i}^{-1}=H_{R(L)}(\phi_{L(R)}),\label{UiSC}
\end{eqnarray}
and the Hamiltonian for NLSM is invariant under the same operation, i.e.,
\begin{eqnarray}
\mathcal{U}_{i}H_{NLSM}({\bf{B}})\mathcal{U}_{i}^{-1}=H_{NLSM}({\bf{B}}).\label{sym1}
\end{eqnarray}
The mirror reflection $\mathcal{M}_{yz}$ in the inversion-rotation joint operation exchanges the phases of the left SC and the right SC and will invert the flow direction of the Josephson current. Then, we have
\begin{eqnarray}
I({\bf{B}},\phi)=-I({\bf{B}},-\phi),\label{IB}
\end{eqnarray}
which is independent of the direction and the magnitude of the external field. The relation $I(\phi)=-I(-\phi)$ in Eq.(\ref{IB}) will limit the generation of JDE for $B\ne0$ and $\lambda=0$.

\subsection{$\lambda\ne0$ and $B=0$}
\begin{figure}[!htb]
\centerline{\includegraphics[width=0.9\columnwidth]{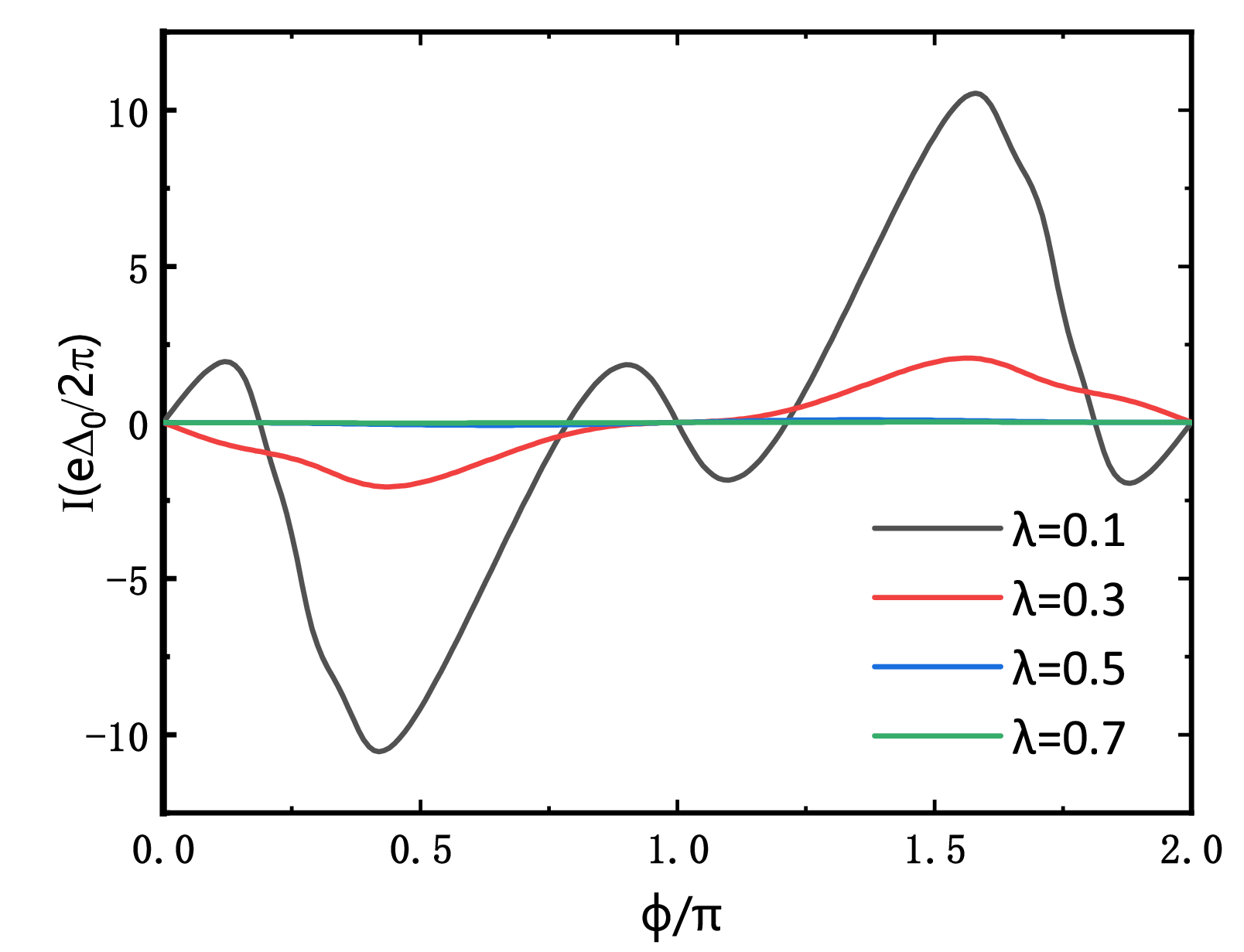}}
\caption{CPRs for different strengths of RSOC with $B=0$.}\label{fig4}
\end{figure}	

In Fig.\ref{fig4}, we give CPRs for different strengths of RSOC. It is found that CPRs in Fig.\ref{fig4} possess the similar behaviors to those in Fig.\ref{fig3}(c). Actually, the presence of RSOC also plays a role in opening a gap in the energy bands of NLSM, which brings the same effect as the field with only the in-plane component in Fig.\ref{fig3}(c). As the strength of RSOC is increased, the critical current of CPRs will change from the finite value to a value close to zero. The switch of the Josephson current can be tuned by the strength of RSOC. Furthermore, we stress that the change of the size of the nodal line in NLSM, i.e., the value of $\mu_M$, has very little impact on the switch effect. In other words, the switch effect shown in Fig.3 and Fig.4 can survive for different values of $\mu_M$. In addition, CPRs in Fig.\ref{fig4} only include contribution of the $\sin{n\phi}$ term as those in Figs. \ref{fig2} and \ref{fig3}. This indicates the absence of JDE in this situation although NLSM itself has broken the time-reversal symmetry and RSOC breaks the inversion symmetry. The absence of JDE here can also be understood from the symmetry analysis. We introduce the joint operation $\mathcal{U}_{tm}=\mathcal{T}\mathcal{M}_{xz}$ with $\mathcal{T}$ being the time-reversal operation\cite{Cheng2}. The action of $U_{tm}$ on the Nambu spinor $(c_{\bf{k}\uparrow},c_{\bf{k}\downarrow},c^{+}_{-\bf{k}\uparrow},c^{+}_{-\bf{k}\downarrow})$ can be expressed as $U_{tm}c_{{\bf{k}\sigma}}U_{tm}^{-1}=-c_{(-k_x,k_y)\sigma}$ and $U_{tm}c^{+}_{{\bf{k}\sigma}}U_{tm}^{-1}=-c^{+}_{(-k_x,k_y)\sigma}$. The transformation of the Hamiltonian for the left (right) SC under the joint operation is given by
\begin{eqnarray}
\mathcal{U}_{tm}H_{L(R)}(\phi_{L(R)})\mathcal{U}_{tm}^{-1}=H_{L(R)}(-\phi_{L(R)}).
\end{eqnarray}
The Hamiltonian for NLSM is invariant under the joint operation, i.e.,
\begin{eqnarray}
\mathcal{U}_{tm}H_{NLSM}(\lambda)\mathcal{U}_{tm}^{-1}=H_{NLSM}(\lambda).\label{sym2}
\end{eqnarray}
Since the time-reversal operation in $\mathcal{U}_{tm}$ will invert the flow direction of the Josephson current, one has the relation $I(\phi)=-I(-\phi)$ for $\lambda\ne0$ and $B=0$. This relation implies the absence of JDE in this situation, which can also be concluded from the isotropic energy bands and Fermi surfaces in Fig.\ref{fig1}(c). %, which can also be derived from the invariance of NLSM under the inversion operation $\mathcal{U}_{i}$, i.e.,
%\begin{eqnarray}
%\mathcal{U}_{i}H_{NLSM}(\lambda)\mathcal{U}_{i}^{-1}=H_{NLSM}(\lambda).
%\end{eqnarray}
%According to the transformation of Hamiltonians for SCs under $\mathcal{U}_{i}$ in Eq.(\ref{UiSC}), one can also obtain the relation %$I(\phi)=-I(-\phi)$ for $\lambda\ne0$ and $B=0$. %

\subsection{$B\ne0$ and $\lambda\ne0$}
\begin{figure}[!htb]
\centerline{\includegraphics[width=1\columnwidth]{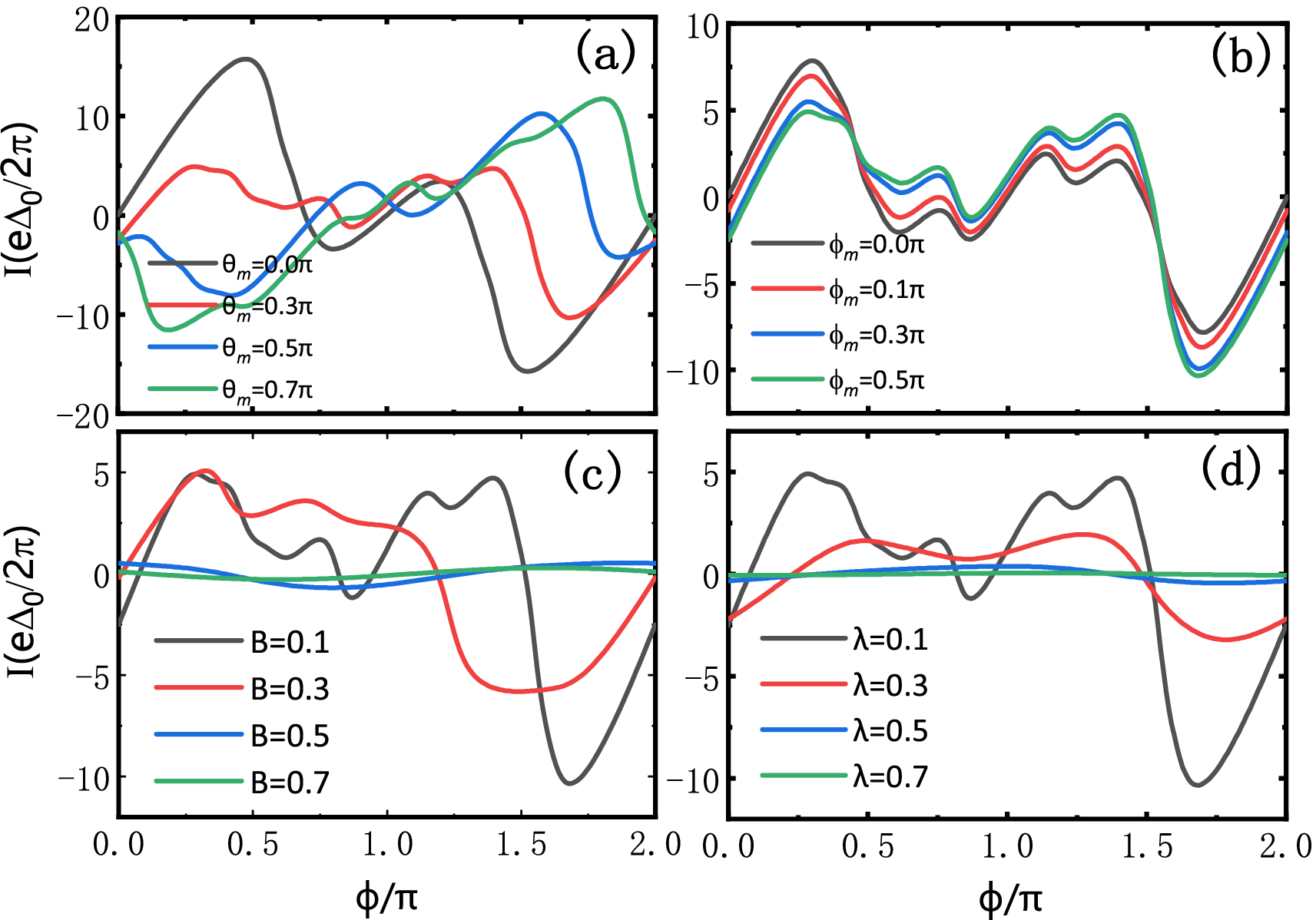}}
\caption{(a) CPRs for different polar angles of the external field with $\phi_m=0.5\pi$, $\lambda=0.1$ and $B=0.1$. (b) CPRs for different azimuthal angles of the external field with $\theta_m=0.3\pi$, $\lambda=0.1$ and $B=0.1$. (c) CPRs for different magnitudes of the external field with $\theta_m=0.3\pi$, $\phi_m=0.5\pi$ and $\lambda=0.1$. (d) CPRs for different strengths of RSOC with $\theta_m=0.3\pi$, $\phi_m=0.5\pi$ and $B=0.1$.}\label{fig5}
\end{figure}	

For $B\ne0$ and $\lambda\ne0$, the relation $I(\phi)=-I(-\phi)$ can be ruined by the coexistence of the external field and RSOC. For the field deviating from the $z$ axis such as $\theta_m=0.3\pi,0.5\pi$ and $0.7\pi$ in Fig.{\ref{fig5}}(a), the $\cos{\phi}$ term will contribute to the Josephson current. The critical value of the current along the $+x$ direction and that along the $-x$ direction are not equal, which is JDE realized in our junctions. For a given value of the polar angle with $\theta_m=0.3\pi$, JDE can also be observed for the nonzero azimuthal angles as shown in Fig.{\ref{fig5}}(b). When $\phi_m$ is increased from $0.1\pi$ to $0.5\pi$, the critical current for the positive direction is suppressed and that for the negative direction is enhanced. As a result, the optimal nonreciprocity of supercurrent can be achieved when the field is located in the $yz$ plane perpendicular to the direction of the Josephson current. The presences of the $\cos{\phi}$-type current and JDE in Figs. \ref{fig5}(a) and \ref{fig5}(b) originate from the simultaneous breaking of the symmetries in Eqs.(\ref{sym1}) and (\ref{sym2}) obeyed by NLSM. The two symmetries are responsible for the establishment of the relation $I(\phi)=-I(-\phi)$. The coexistence of the external field and RSOC can break these symmetries simultaneously and bring the contribution of $\cos{\phi}$-type current and JDE in our junctions. However, for the specific orientation of the field with $B_y=0$, the $\cos{\phi}$ contribution and JDE will vanish as shown in Fig.{\ref{fig5}}(a) with $\theta_m=0$ and in Fig.{\ref{fig5}}(b) with $\phi_m=0$. In this situation, NLSM still preserves the symmetry under the $\mathcal{U}_{tm}$ operation although the inversion-rotation symmetry is broken. The relation $I(\phi)=-I(-\phi)$ is protected by the symmetry under the joint operation of the time-reversal and the mirror reflection about the $xz$ plane. Therefore, the $\cos{\phi}$ term is absent in CPRs and JDE does not emerge. In other words, the coexistence of the external field and RSOC is the necessary but not the sufficient condition. The emergence of JDE also requires $B_y\ne0$, which is similar to the property of the Josephson junctions based on the topological insulator\cite{Lu}. Actually, under the condition of $\lambda\ne0$ and $B_y=0$, the energy bands and Fermi surfaces are either isotropic as shown in Fig.\ref{fig1}(c) or symmetric about the $k_y$ axis as shown in Fig.\ref{fig1}(d). Only under the condition of $\lambda\ne0$ and $B_y\ne0$, the energy bands and Fermi surfaces are neither isotropic nor symmetric about the $k_y$ axis as shown in Fig.\ref{fig1}(e), which will bring JDE\cite{Yuan}.

In Fig.{\ref{fig5}}(c) and Fig.{\ref{fig5}}(d), we present CPRs for different magnitudes of the external field and for different strengths of RSOC, respectively. The polar angle and the azimuthal angle of the field are taken as $\theta_m=0.3\pi$ and $\phi_m=0.5\pi$, which satisfy the condition for the realization of JDE as discussed above. It is found that the nonreciprocity of the Josephson current exhibits the similar character under variations of the field and RSOC. When the magnitude of the field or the strength of RSOC is reduced, the nonreciprocity of the Josephson current is enhanced. This is an advantage of JDE realized in our junctions. In experiment, a small RSOC is usually more easy to be introduced in NLSM and a small field is beneficial for the practical applications.

\begin{figure}[!htb]
\centerline{\includegraphics[width=0.9\columnwidth]{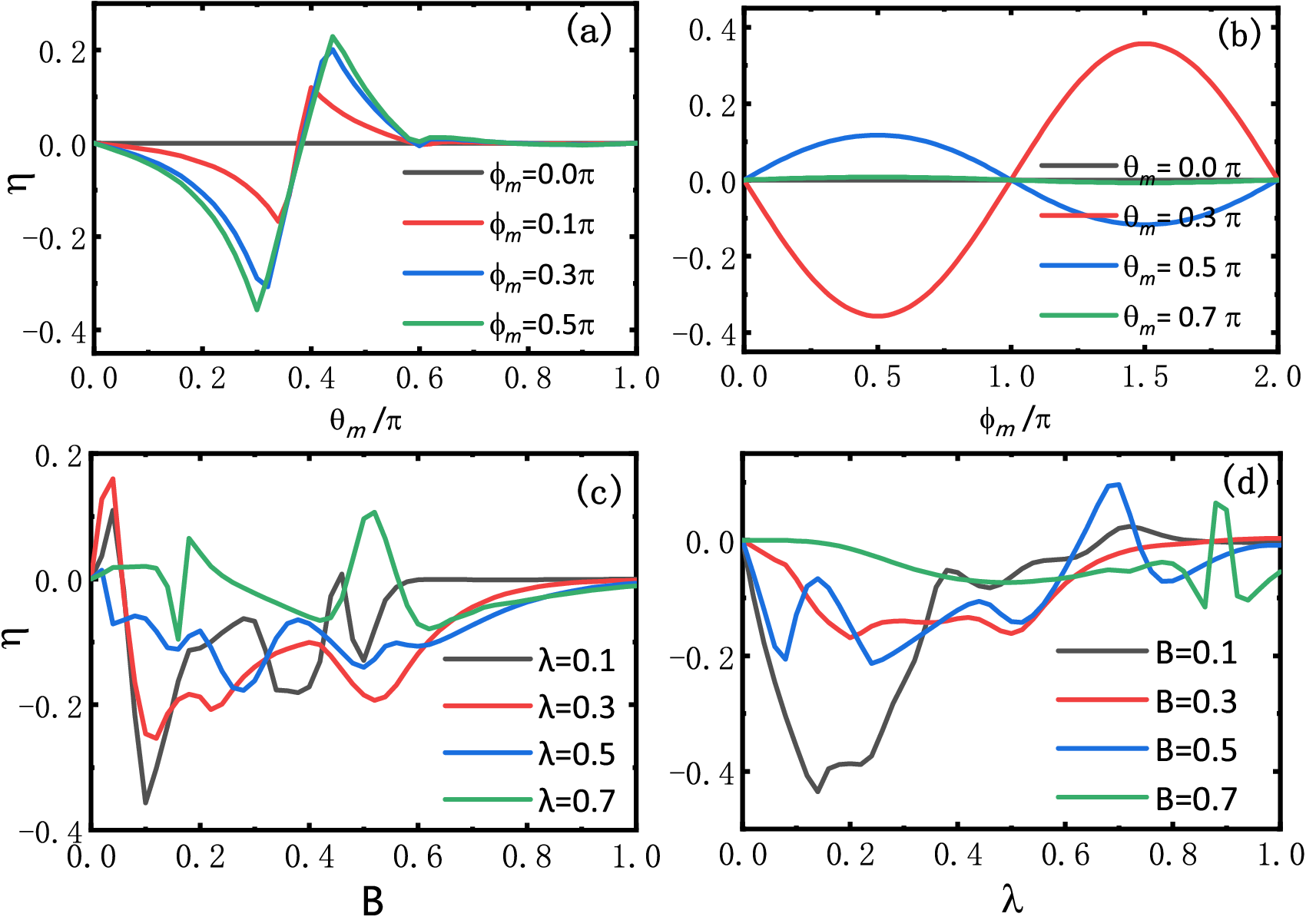}}
\caption{The variations of the diode efficiency $\eta$ as (a) the polar angle, (b) the azimuthal angle, (c) the magnitude of the field and (d) the strength of RSOC. Other parameters are taken as $B=0.1$ and $\lambda=0.1$ in (a) and (b) and $\theta_m=0.3\pi$ and $\phi_m=0.5\pi$ in (c) and (d).}\label{fig6}
\end{figure}		

In order to quantitatively measure the nonreciprocity of the Josephson current in our junctions, we define the diode efficiency as
\begin{eqnarray}
\eta=\frac{I_{c+}-I_{c-}}{I_{c+}+I_{c-}},
\end{eqnarray}
with $I_{c+}=\text{max}[I(0<\phi<2\pi)]$ and $I_{c-}=\text{max}[-I(0<\phi<2\pi)]$ being the critical value for the positive direction and that for the negative direction, respectively. In Fig.\ref{fig6}, we present the variations of diode efficiency $\eta$ as parameters. For $B=0.1$ and $\lambda=0.1$ in Figs. \ref{fig6}(a) and \ref{fig6}(b), the diode efficiency is exactly zero for $\phi_m=0$ or $\theta_m=0$. This is consistent with CPRs for $\phi_m=0$ and $\theta_m=0$ in Figs. \ref{fig5}(a) and \ref{fig5}(b) discussed above. For $\phi_m\ne0$, the diode efficiency experiences one obvious oscillation as $\theta_m$ and then is obviously suppressed as shown in Fig.\ref{fig6}(a). The extremum of the diode efficiency can be obtained around $\theta_m=0.4\pi$ and its absolute value can exceed $30\%$ for $\phi_m=0.5\pi$. For $\theta_m\ne0$, the diode efficiency is symmetric about $\phi_m=0.5\pi$ ($\phi_m=1.5\pi$) in the angle range $0<\phi_m<\pi$ ($\pi<\phi_m<2\pi$) while it is antisymmetric about $\phi_m=\pi$ as shown in Fig.\ref{fig6}(b). The extremum of the diode efficiency is obtained at $\phi_m=0.5\pi$ or $\phi_m=1.5\pi$ for different $\theta_m$. This is also consistent with CPRs in Fig.\ref{fig5}(b) discussed above, where the nonreciprocity of the Josephson current is enhanced when $\theta_m$ is increased from $\theta_m=0.1\pi$ to $\theta_m=0.5\pi$.

In Fig.6(a), a sign change occurs at $\theta_m\approx0.4\pi$, which is related to the $0$-$\pi$-like transition\cite{Costa}. Now, we consider the $\eta$ curve for $\phi_m=0.5\pi$, i.e., the green curve in Fig.6(a). Its corresponding CPRs for different values of $\theta_m$ are presented in Fig.5(a). As shown in Fig.5(a), CPRs for $\theta_m=0$ and $\theta_m=0.3\pi$ are more like the $\sin\phi$-type function while CPRs for $\theta_m=0.5\pi$ and $0.7\pi$ are more like the $-\sin{\phi}$-type function. As an integral of $\phi$, the free energy for the former CPRs correspond to the $0$-like junctions with the minimum of the free energy near $\phi=0$ and the latter CPRs correspond to the $\pi$-like junctions with the minimum of the free energy near $\phi=\pi$. When $\theta_m$ is increased from $0$ to $0.7\pi$, the $0$-$\pi$-like transition will happen at $\theta_m\approx0.4\pi$, which shows itself as the sign change of the efficiency. The diode efficiencies for $\phi_m=0.1\pi$ and $\phi_m=0.3\pi$ possess the similar behavior as shown in Fig.6(a).
%Their corresponding CPRs for different values of $\theta_m$ are not presented here.

For $\theta_m=0.3\pi$ and $\phi_m=0.5\pi$, the variations of $\eta$ as the magnitude of the external field and the strength of RSOC are presented in Figs. \ref{fig6}(c) and \ref{fig6}(d). First, the quantity $\eta$ shows the oscillatory behavior, which is a typical character of the diode efficiency in various Josephson structures\cite{Tanaka,Kokkeler,Costa,Banerjee}. Second, the quantity $\eta$ can reach its maximum absolute value for small field and small RSOC. In Fig.\ref{fig6}(c), the absolute value of $\eta$ reaches $43\%$ at $\lambda=0.14$ for $B=0.1$. Even at $\lambda=0.06$, the absolute value exceeding $20\%$ can be obtained by $\eta$. In Fig.\ref{fig6}(d), the absolute value of $\eta$ reaches $35.6\%$ at $B=0.1$ for $\lambda=0.1$. Even at $B=0.04$, the value exceeding $10\%$ can be obtained by $\eta$. In contrast, when the magnitude of the field in Fig.\ref{fig6}(c) or the strength of RSOC in Fig.\ref{fig6}(d) is increased, the diode efficiency shows a decreasing trend. This feature of $\eta$ is consistent with CPRs in Figs.5(c) and (d). For a given value of $\lambda$, the increase of $B$ will enlarge the gap in the energy bands of NLSM and the Josephson current will be significantly weakened as shown in Fig.5(c). The difference of $I_{C+}$ and $I_{C-}$ will decrease accordingly. For a given value of $B$, the increase of $\lambda$ will lead to the similar behavior as shown in Fig.5(d). Therefore, the efficiency of JDE is suppressed when $B$ or $\lambda$ is increased as shown in Figs.6(c) and (d). This is the reason that the larger nonreciprocity in our junctions is obtained for small values of $B$ and $\lambda$.
The high diode efficiency for small field and small RSOC in our junctions improves the experimental feasibility of JDE.

Finally, we discuss the realistic scales of parameters for JDE in our junctions. In our calculations, $\mu_M=2$ has been taken and the Zeeman energy is denoted by $B$. If we take $B=0.04$ for the diode efficiency exceeding $10\%$, the Zeeman energy will be $0.02\mu_M$. For NLSM, the order of magnitude of $\mu_M$ can be $1meV$. Actually, its value can be tuned by strain\cite{Jin}. In this situation, the Zeeman energy corresponds to an external magnetic field of the value about $300mT$. In experiment, the order of magnitude of the field for observing the superconducting diode effect is several hundred millitesla or even several tesla\cite{Baumgartner,Ando}. On the other hand, the efficiency about $4\%$ can be observed in experiment\cite{Ando}, which means that smaller field and larger $\mu_M$ are still possible for the observable JDE in our junctions. For RSOC, $\lambda=0.1$ means $\frac{\lambda}{a}=0.05\mu_M$. The magnitude of RSOC can be solved as $5\times10^{-14}eVm$ if $a=1nm$. This value is far less than the order of magnitude of $10^{-11}$ for RSOC in heterostructures such as InAlAs/InGaAs\cite{Nitta}. Since the magnitude of RSOC can be well tuned by applying a voltage\cite{Nitta,Shalom}, the value of RSOC for JDE in our junctions is achievable. Other parameters are taken as $t_0=0.5\mu_Ma^2$, $t=0.5\mu_M$, $\mu=\mu_M$ and $\Delta_0=0.005\mu_{M}$. In addition, we don't consider the suppression of superconductivity in SCs under the magnetic field. According to the Ginzburg-Landau theory, the superconducting gap can be expressed as $\Delta_0=\tilde{\Delta}_{0}\sqrt{1-\frac{B}{B_c}}$ with $B_{c}$ being the critical field and $\tilde{\Delta}_{0}$ being the gap for $B=0$. From the expression, one can find that $\Delta_0  \approx \tilde{\Delta}_0$ still holds when $B<\frac{B_c}{4}$. In our paper, $\Delta_0$ denotes the gap magnitude under a given field $B$. It can be taken as the unit of other parameters and the numerical results are not affected in this situation.

\section{\label{sec4}Conclusions}
The Josephson effects in the SC/NLSM/SC junctions are systematically studied using the Green's function method. If only an external field is exerted on NLSM or only RSOC is introduced in NLSM, CPRs in the junctions will strongly depend on the direction and the magnitude of the field or the strength of RSOC. The $0$-$\pi$ transition can be achieved by adjusting the direction of the field, the magnitude of the field or the strength of RSOC. Simultaneously, the change of the direction and the magnitude of the field or the strength of RSOC can lead to the switch effect of the Josephson current. If an external field and RSOC coexist in NLSM, the symmetries protecting the reciprocity of the Josephson current will be broken and JDE with the nonreciprocity of supercurrent will be realized. The diode efficiency can reach a value exceeding $40\%$ and can also be regulated by rotating the field and tuning the magnitude of the field or the strength of RSOC. The controllable JDE, $0$-$\pi$ transition and switch effect found in our junctions provide possibilities for the design of quantum devices based on NLSM.

\section*{\label{sec5}ACKNOWLEDGMENTS}
This work was financially supported
by the National Natural Science Foundation of China
under Grants Nos. 12474046 and 12374034,
the National Key R and D Program of China (Grant No. 2024YFA1409002),
the Innovation Program for Quantum Science and Technology (2021ZD0302403),
and the projects ZR2023MA005 and ZR2022QA110 supported by Shandong Provincial Natural Science Foundation. We acknowledge
the High-performance Computing Platform of Peking University
for providing computational resources.

\section*{\label{sec5} APPENDIX}
\setcounter{equation}{0}
\setcounter{figure}{0}
\renewcommand{\theequation}{A\arabic{equation}}
\renewcommand{\thefigure}{A\arabic{figure}}

\subsection{The Hamiltonian and electronic structure of NLSM}
The Hamiltonian of NLSM in the spin space can be written as
\begin{eqnarray}
h_{M}=\sum_{\bf{k}}\psi_{M}^{+}({\bf{k}})\hat{h}_{M}\psi_{M}({\bf{k}}),\label{hnlsm}
\end{eqnarray}
with the single-particle Hamiltonian
\begin{eqnarray}
\hat{h}_{M}({\bf{k}})=\epsilon_M\sigma_{z}+{\bf{B}}\cdot{\bf{\sigma}}+\lambda({\bf{\sigma}}\times{\bf{k}})\cdot\hat{z},\label{sh}
\end{eqnarray}
and $\psi_{M}({\bf{k}})=(c_{M{\bf{k}}\uparrow},c_{M{\bf{k}}\downarrow})^T$.
The first term in Eq.({\ref{sh}}) describes the energy of electrons in the intrinsic NLSM\cite{Cheng4,Jin}. The second is the Zeeman term describing the coupling between the external magnetic field and spin of electrons. The third is RSOC with $\hat{z}$ being the unit vector along the out-of-plane direction. Here, ${\bf{\sigma}}=(\sigma_x,\sigma_y,\sigma_z)$ is the Pauli matrix in the spin space.

From the Hamiltonian in Eq.(\ref{hnlsm}), one can obtain the BdG Hamiltonian for NLSM in the particle-hole$\otimes$spin space. The BdG Hamiltonian can be written as
\begin{eqnarray}
H_{M}=\sum_{\bf{k}} \Psi_{M}^{+}({\bf{k}}) \check{H}_{M}({\bf{k}})\Psi_{M}({\bf{k}}),
\end{eqnarray}
with
\begin{eqnarray}
\check{H}_{M}({\bf{k}})=\left(\begin{array}{cc}
\hat{h}_{M}({\bf{k}})&0\\
0&-\hat{h}^{*}_{M}(-{\bf{k}})\end{array}\right).
\end{eqnarray}
Here, $-\hat{h}^{*}_{M}(-{\bf{k}})$ denotes the Hamiltonian in the spin space for holes. $\check{H}_{M}({\bf{k}})$ can be rewritten as more compact form in Eq.(\ref{NLSMH}) with the direct product of the Pauli matrices in the spin and particle-hole spaces.

The energy bands of electrons can be solved through diagonalizing the single-particle Hamiltonian in Eq.(\ref{sh}), which are given by
\begin{eqnarray}
E_{\pm}=\pm\sqrt{(\epsilon_M+B_{z})^2+B_{\parallel}^2+\lambda^2k^2+2\lambda({\bf{B}}\times{\bf{k}})\cdot\hat{z}},\label{eb}
\end{eqnarray}
with ${\bf{B}}_{\parallel}=(B_{x},B_{y})$ being the in-plane component of the external field ${\bf{B}}$.

Based on Eq.(\ref{eb}), the energy bands of electrons in NLSM have the following features.

(1) For $\lambda=0$ and $B_{\parallel}=0$, there is no gap in the energy bands and the Fermi surfaces are isotropic as shown in Fig.\ref{fig1}(b). Even for $B_{z}\ne0$, its effect will be absorbed by $\mu_M$ in $\epsilon_M$ and will not open a gap. In this situation, spin is still a good number. For clarity, we have used different colors to distinguish energy bands and Fermi surfaces for different spins in Fig.\ref{fig1}(b).

(2) For $\lambda=0$ and $B_{\parallel}\ne0$ or $\lambda\ne0$ and $B_{\parallel}=0$, a gap will open in the energy bands. However, the Fermi surfaces are still isotropic as shown in Fig.{\ref{fig1}}(c).

(3) For $\lambda\ne0$ and $B_{\parallel}\ne0$, the Fermi surface will not be isotropic as shown in Figs.{\ref{1}}(d) and (e). If $B_{x}\ne0$ and $B_{y}=0$, the Fermi surfaces are symmetric about the $k_y$ axis as shown in Fig.{\ref{fig1}}(d). If $B_x=0$ and $B_y\ne0$, the Fermi surfaces are symmetric about the $k_x$ axis as shown in Fig.{\ref{fig1}}(e), which will bring JDE in our junctions as discussed in the main text.

\subsection{The discretization of the Continuous Hamiltonian.}\label{A}
\begin{figure}[!htb]
\centerline{\includegraphics[width=1.0\columnwidth]{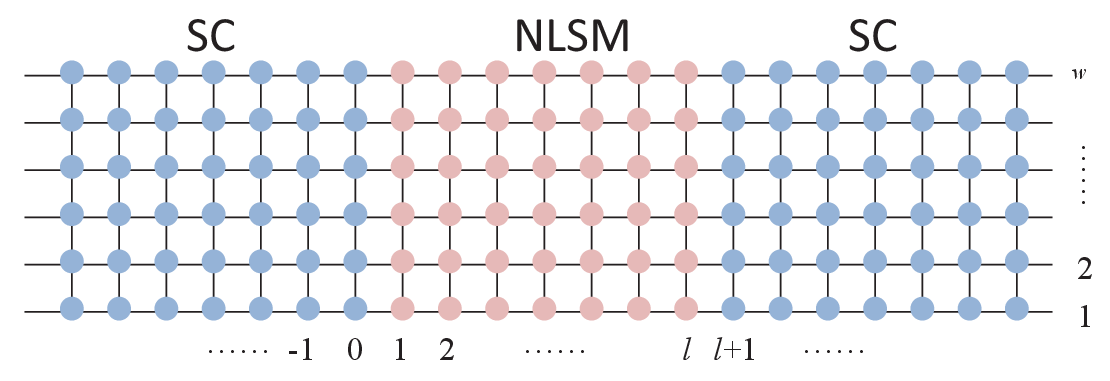}}
\caption{The two-dimensional square lattice on which the continuum Hamiltonian is discretized.}
\label{figA1}
\end{figure}	
We use the finite difference method to discretize the continuous Hamiltonian on a two-dimensional square lattice with the lattice constant $a$\cite{Cheng2,Cheng5,Cheng6}. For the left (right) SC, the onsite Hamiltonian matrix $\check{H}_{L(R)0}$ in Eq.(\ref{10}) is given by
\begin{eqnarray}
\check{H}_{L(R)0}&=-\Delta_{0}[\cos{\phi_{l(r)}}\tau_{y}+\sin{\phi_{l(r)}}\tau_{x}]\otimes\sigma_{y}\nonumber\\
&+e_{L(R)}\tau_z\otimes\sigma_0,\label{A1}
\end{eqnarray}
with $e_{L(R)}=\frac{4t_0}{a^2}-\mu_{L(R)}$.
The Hamiltonian matrices $\check{H}_{{L(R)}x}$ and $\check{H}_{{L(R)}y}$ for the hopping from a lattice point to its neighbor point along the $x$ direction and the $y$ direction are given by
\begin{eqnarray}
\check{H}_{{L(R)}x}&=&-\frac{t_0}{a^2}\tau_z\otimes\sigma_0.\label{A2}\\
\check{H}_{{L(R)}y}&=&-\frac{t_0}{a^2}\tau_z\otimes\sigma_0,\label{A3}
\end{eqnarray}

For NLSM, the onsite Hamiltonian matrix $\check{H}_{M}$ in Eq.(\ref{11}) is given by
\begin{eqnarray}
\check{H}_{M}&=B_{x}\tau_{z}\otimes\sigma_{x}+B_{y}\tau_{0}\otimes\sigma_{y}+B_{z}\tau_{z}\otimes\sigma_{z}\nonumber\\
&+e_{M}\tau_{z}\otimes\sigma_{z},\label{A4}
\end{eqnarray}
with $e_{M}=\frac{4t_0}{a^2}-\mu_{M}$. The Hamiltonian matrices $\check{H}_{{M}x}$ and $\check{H}_{{M}y}$ signifying the hopping from a lattice point to its neighbor point along the $x$ direction and the $y$ direction are given by
\begin{eqnarray}
\check{H}_{{M}x} &=&-\frac{t_0}{a^2}\tau_z\otimes\sigma_z+\frac{\lambda}{2a}\tau_z\otimes i\sigma_y,\label{A5}\\
\check{H}_{{M}y}&=&-\frac{t_0}{a^2}\tau_z\otimes\sigma_z+\frac{i\lambda}{2a}\tau_x\otimes \sigma_0.\label{A6}
\end{eqnarray}
The above Hamiltonian matrices in Eqs.(\ref{A1})-(\ref{A6}) are just the ones in the discrete Hamiltonians in Eqs.(\ref{10}) and (\ref{11}). Using these matrices, one can obtain the matrices $H_{L}^{11}$, $H_{L}^{12}$, $H_{L}^{21}$, $H_M^{11}$, $H_{M}^{12}$, $H_{M}^{21}$, $H_{R}^{11}$, $H_{R}^{12}$ and $H_{R}^{21}$ for the calculations of Green's functions\cite{Cheng2,Cheng5,Cheng6}.

\setcounter{equation}{0}
\setcounter{figure}{0}
\renewcommand{\theequation}{B\arabic{equation}}
\renewcommand{\thefigure}{B\arabic{figure}}
\subsection{The derivation of the Green's functions} \label{B}
In order to derive the surface Green's function of the left SC, i.e., the Green's function for the rightmost column, we construct the M$\ddot{\text{o}}$bius transformation matrix as\cite{Cheng2,Cheng5,Cheng6}
\begin{eqnarray}
X_{L}=\left(\begin{array}{cc}
0&(H_{L}^{12})^{-1}\\
-(H_{L}^{12})^{\dagger}&[(E+i\gamma)-H_{L}^{11}] (H_{L}^{12})^{-1}
\end{array}\right),
\end{eqnarray}
with a small positive quantity $\gamma$. The matrix $X_{L}$ can be diagonalized by the matrix $U_{L}$ which satisfies
\begin{eqnarray}	U_{L}^{-1}X_{L}U_{L}=\text{diag}(\lambda_{L1},\lambda_{L2},\lambda_{L3}, \cdots).\label{B2}
\end{eqnarray}
Here, the eigenvalues $\lambda_{L}$ are ordered from left to right according to the increasing magnitude of their absolute values. If $U_{L}$ is expressed as
\begin{eqnarray}
U_{L}=\left(\begin{array}{cc}
U_{L}^{11}&U_{L}^{12}\\\\
U_{L}^{21}&U_{L}^{22}
\end{array}\right),\label{B3}
\end{eqnarray}
the surface Green's function for the left SC will be given by
$g_{L}^{r}=U_{L}^{12}(U_{L}^{22})^{-1}$. Through the similar process, the surface Green's function of the right SC, i.e., the Green's function for the leftmost column, can be given by $g_{R}^{r}=U_{R}^{12}(U_{R}^{22})^{-1}$.

Next, to calculate the retarded Green's function for the leftmost column of middle NLSM, we will use the recursive algorithm.
When recursion is carried out from right to left, the corresponding Green's function is denoted by the superscript $R$ and the retarded Green's function is denoted by superscript $r$. The Green's function for the rightmost column of NLSM is given by
\begin{eqnarray}
\mathcal{G}_{M}^{Rr}(E,l)=[E-H_{M}^{11}-\tilde{T}_{R}^{\dagger}g_{R}^{r}(E)\tilde{T}_{R}]^{-1}.\label{B4}
\end{eqnarray}
with $\tilde{T}_{R}=1_{w\times w}\otimes \check{T}_{R}$. The Green's function for the jth column can be written as
\begin{eqnarray}
\mathcal{G}_{M}^{Rr}(E,j)=&[E-H_{M}^{11}-H_{M}^{12}\mathcal{G}_{M}^{Rr}(E,j+1)H_{M}^{21}]^{-1}.\nonumber\\
&&\label{B5}
\end{eqnarray}
By continuing the recursion, the retarded Green's function for the leftmost column of NLSM can be expressed as
\begin{equation}
\begin{split}
\mathcal{G}_{M}^{r}(E,1)&=[E-H_{M}^{11}-\tilde{T}_{L}^{\dagger}g_{L}^{r}(E)\tilde{T}_{L}\\
&-H_{M}^{12}\mathcal{G}_{M}^{Rr}(E,2)H_{M}^{21}]^{-1}.
\end{split}
\end{equation}
The corresponding advanced Green's function is given by $\mathcal{G}_{M}^{a}(E)=[\mathcal{G}_{M}^{r}(E)]^{\dagger}$. Using the fluctuation-dissipation theorem, the lesser Green's function in the expression of the Josephson current in Eq.(\ref{16}) can be expressed as
\begin{eqnarray}
G_{ML}^{<}(E)=-f(E)[G_{ML}^{r}(E)-G_{ML}^{a}(E)].
\end{eqnarray}	


\section*{REFERENCES}
\begin{thebibliography}{}
\bibitem{Wu}
H. Wu, Y. Wang, Y. Xu, P. K. Sivakumar, C. Pasco, U. Filippozzi, S. S. P. Parkin, Y.-J. Zeng, T. McQueen, and M. N. Ali, The field-free Josephson diode in a van der Waals heterostructure, Nature, \textbf{604}, 653 (2022).
\bibitem{Pal}
B. Pal, A. Chakraborty, P. K. Sivakumar, M. Davydova, A. K. Gopi, A. K. Pandeya, J. A. Krieger, Y. Zhang, M. Date, S. Ju, N. Yuan, N. B. M. Schr$\ddot{\text{o}}$ter, L. Fu, and S. S. P. Parkin, Josephson diode effect from Cooper pair momentum in a topological semimetal, Nature, \textbf{18}, 1228 (2022).
\bibitem{Baumgartner}
C. Baumgartner, L. Fuchs, A. Costa, S. Reinhardt, S. Gronin, G. C. Gardner, T. Lindemann, M. J. Manfra, P. E. Faria, Jr., D. Kochan, J. Fabian, N. Paradiso, and C. Strunk, Supercurrent rectification and magnetochiral effects in symmetric Josephson junctions, Nat. Nanotechnol. \textbf{17}, 39 (2022).
\bibitem{Banerjee}
A. Banerjee, M. Geier, M. A. Rahman, C. Thomas, T. Wang, M. J. Manfra, K. Flensberg, and C. M. Marcus, Phase Asymmetry of Andreev Spectra from Cooper-Pair Momentum, Phys. Rev. Lett. \textbf{131}, 196301 (2023).
\bibitem{Lin}
J.-X. Lin, P. Siriviboon, H. D. Scammell, S. Liu, D. Rhodes, K. Watanabe, T. Taniguchi, J. Hone, M. S. Scheurer, and J. I. A. Li, Zero-field superconducting diode effect in small-twist-angle
trilayer graphene, Nature, \textbf{18}, 1221 (2022).
\bibitem{Trahms}
M. Trahms, L. Melischek, J. F. Steiner, B. Mahendru, I. Tamir, N. Bogdanoff, O. Peters, G. Reecht, C. B. Winkelmann, F. von Oppen, and K. J. Franke, Diode effect in Josephson junctions with a
single magnetic atom, Nature, \textbf{615}, 628 (2023).
\bibitem{Matsuo}
S. Matsuo, T. Imoto, T. Yokoyama, Y. Sato, T. Lindemann, S. Gronin, G. C. Gardner, M. J. Manfra, and S. Tarucha, Josephson diode effect derived from short-range coherent coupling, Nature, \textbf{19}, 1636 (2023).
\bibitem{Halterman}
K. Halterman,M. Alidoust, R. Smith, and S. Starr, Supercurrent diode effect, spin torques, and robust zero-energy peak in planar half-metallic trilayers, Phys. Rev. B 105, 104508 (2022).
\bibitem{Davydova}
M. Davydova, S. Prembabu, and L. Fu, Universal Josephson diode effect, Sci. Adv. 8, eabo0309 (2022).
\bibitem{Fominov}
Ya. V. Fominov and D. S. Mikhailov, Asymmetric higherharmonic SQUID as a Josephson diode, Phys. Rev. B 106, 134514 (2022).
\bibitem{YZhang}
Y. Zhang, Y. Gu, P. Li, J. Hu, and K. Jiang, General theory of Josephson diodes, Phys. Rev. X 12, 041013 (2022).
\bibitem{Souto}
R. S. Souto, M. Leijnse, and C. Schrade, Josephson diode effect in supercurrent interferometers, Phys. Rev. Lett. 129, 267702 (2022).
\bibitem{Tanaka}
Y. Tanaka, B. Lu, and N. Nagaosa, Theory of giant diode effect in d-wave superconductor junctions on the surface of a topological insulator, Phys. Rev. B 106, 214524 (2022).
\bibitem{Kokkeler}
T. H. Kokkeler, A. A. Golubov, and F. S. Bergeret, Field-free anomalous junction and superconducting diode effect in spinsplit
superconductor/topological insulator junctions, Phys. Rev. B 106, 214504 (2022).
\bibitem{Cheng1}
Q. Cheng and Q.-F. Sun, Josephson diode based on conventional superconductors and a chiral quantum dot, Phys. Rev. B 107, 184511 (2023).
\bibitem{YFSun1}
Y.-F. Sun, Y. Mao, and Q.-F. Sun, Design of Josephson diode based on magnetic impurity, Phys. Rev. B 108, 214519 (2023).
\bibitem{Cheng2}
Q. Cheng, Y. Mao, and Q.-F. Sun, Field-free Josephson diode effect in altermagnet/normal metal/altermagnet junctions, Phys. Rev. B \textbf{110}, 014518 (2024).
\bibitem{YFSun2}
Y.-F. Sun, Y. Mao, and Q.-F. Sun, Design of a Josephson diode based on double magnetic impurities, Phys. Rev. B \textbf{111}, 054515 (2025).
\bibitem{Lu}
B. Lu, S. Ikegaya, P. Burset, Y. Tanaka, and N. Nagaosa, Tunable Josephson diode effect on the surface of topological insulators, Phys. Rev. Lett. 131, 096001 (2023).
\bibitem{Maoadd}
Y. Mao, Q. Yan, Y.-C. Zhuang, and Q.-F. Sun, Universal Spin Superconducting Diode Effect from Spin-Orbit Coupling, Phys. Rev. Lett. \textbf{132}, 216001 (2024).
\bibitem{Costa}
A. Costa, J. Fabian, and D. Kochan, Microscopic study of the Josephson supercurrent diode effect in Josephson junctions based on two-dimensional electron gas, Phys. Rev. B 108, 054522 (2023).
\bibitem{Cayao}
J. Cayao, N. Nagaosa, and Y. Tanaka, Enhancing the Josephson diode effect with Majorana bound states, Phys. Rev. B \textbf{109}, L081405 (2024).
\bibitem{Maiani}
A. Maiani, K. Flensberg, M. Leijnse, C. Schrade, S. Vaitiekenas, and R. S. Souto, Nonsinusoidal current-phase relations in semiconductor-superconductor-ferromagnetic insulator devices, Phys. Rev. B 107, 245415 (2023).
\bibitem{Vakili}
H. Vakili, M. Ali, and A. A. Kovalev, Field-free Josephson diode effect in a d-wave superconductor heterostructure, Phys. Rev. B \textbf{110}, 104518 (2024).
\bibitem{Seleznev}
G. S. Seleznev and Ya. V. Fominov, Influence of capacitance and thermal fluctuations on the Josephson
diode effect in asymmetric higher-harmonic SQUIDs, Phys. Rev. B \textbf{110}, 104508 (2024).
\bibitem{Zazunov}
A. Zazunov, J. Rech, T. Jonckheere, B. Gr$\acute{\text{e}}$maud, T. Martin, and R. Egger, Nonreciprocal charge transport and subharmonic structure in voltage-biased Josephson diodes, Phys. Rev. B \textbf{109}, 024504 (2024).
\bibitem{Legg}
H. F. Legg, K. Laubscher, D. Loss, and J. Klinovaja, Parity-protected superconducting diode effect in topological Josephson junctions, Phys. Rev. B \textbf{108}, 214520 (2023).
\bibitem{Ding}
Z. Ding, D. Wang, M. Li, Y. Tao, and J. Wang, Spin-resolved and charge Josephson diode effects in $\alpha-T_{3}$ lattice junctions, Phys. Rev. B \textbf{110}, 155405 (2024).
\bibitem{Patil}
S. Patil, G. Tang, and W. Belzig, Spin-split Andreev bound states and diode effect in an Ising superconductor Josephson junction, Phys. Rev. B \textbf{111}, L060502 (2025).
\bibitem{Ilic}
S. Ili$\acute{\text{c}}$, P. Vertanen, D. Crawford, T. T. Heikkil$\ddot{\text{a}}$, and F. S. Bergeret, Superconducting diode effect in diffusive superconductors and Josephson junctions with Rashba spin-orbit coupling, Phys. Rev. B \textbf{110}, L140501 (2024).
\bibitem{Scharf}
B. Scharf, D. Kochan, and A. Matos-Abiague, Superconducting diode effect in quantum spin Hall insulator based Josephson junctions, Phys. Rev. B \textbf{110}, 134511 (2024).
\bibitem{JXHu}
J.-X. Hu, Z.-T. Sun, Y.-M. Xie, and K. T. Law, Josephson Diode Effect Induced by Valley Polarization in Twisted Bilayer Graphene, Phys. Rev. Lett. \textbf{130}, 266003 (2023).
\bibitem{Cheng3}
Q. Cheng, Z. Hou, and Q.-F. Sun, Double Andreev reflections and double normal reflections in nodal-line semimetal-superconductor junctions, Phys. Rev. B \textbf{101}, 094508 (2020).
\bibitem{YXWang}
Y.-X. Wang, X. Wang, and Y.-X. Li, Double local and double nonlocal Andreev reflections in nodal-line
semimetal-superconducting heterostructures, Phys. Rev. B \textbf{105}, 195402 (2022).
\bibitem{XWang}
X. Wang and W. Luo, Anomalous crossed Andreev reflection in nodal line semimetals, Phys. Rev. B \textbf{109}, 165122 (2024).
\bibitem{Cheng4}
Q. Cheng and Q.-F. Sun, Specular Andreev reflection and its detection, Phys. Rev. B \textbf{103}, 144518 (2021).
\bibitem{YWang}
Y. Wang, H. F. Legg, T. B$\ddot{\text{o}}$merich, J. Park, S. Biesenkamp, A. A. Taskin, M. Braden, A. Rosch, and Y. Ando, Gigantic Magnetochiral Anisotropy in the Topological Semimetal ZrTe$_5$, Phys. Rev. Lett. \textbf{128}, 176602 (2022).
\bibitem{Feng}
B. Feng, R.-W. Zhang, Y. Feng, B. Fu, S. Wu, K. Miyamoto, S. He, L. Chen, K. Wu, K. Shimada, T. Okuda, and Y. Yao, Discovery of Weyl nodal lines in a single-layer ferromagnet, Phys. Rev. Lett. 123, 116401 (2019).
\bibitem{Nie}
S. Nie, H. Weng, and F. B. Prinz,Topological nodal-line semimetals in ferromagnetic rare-earth-metal monohalides, Phys. Rev. B 99, 035125 (2019).
\bibitem{Jin}
L. Jin, X. Zhang, Y. Liu, X. Dai, X. Shen, L. Wang, and G. Liu,Two-dimensional Weyl nodal-line semimetal in a $d^0$ ferromagnetic $K_{2}N$ monolayer with a high Curie temperature, Phys. Rev. B 102, 125118 (2020).
\bibitem{Niu}
C. Niu, P. M. Buhl, H. Zhang, G. Bihlmayer, D. Wortmann, S. Bl$\ddot{\text{u}}$gel, and Y. Mokrousov, Topological Nodal-line Semimetals in Two Dimensions with time-reversal symmetry breaking, arXiv:1703.05540.
\bibitem{Cheng5}
Q. Cheng, Q. Yan, and Q.-F. Sun, Spin-triplet superconductor$-$quantum anomalous Hall insulator$-$spin-triplet superconductor Josephson junctions: $0$-$\pi$ transition, $\phi_0$ phase, and switching effects, Phys. Rev. B \textbf{104}, 134514 (2021).
\bibitem{Cheng6}
Q. Cheng and Q.-F. Sun, Orientation-dependent Josephson effect in spin-singlet superconductor/altermagnet/spin-triplet superconductor junctions, Phys. Rev. B \textbf{109}, 024517 (2024).
\bibitem{Buzdin}
A. I. Buzdin, L. N. Bulaevskii, and S. V. Panyukov, Critical-current oscillations as a function of the exchange field and thickness of the ferromagnetic metal (F) in a S-F-S Josephson junction, JETP Lett. \textbf{35}, 178 (1982).
\bibitem{Ryazanov}
V. V. Ryazanov, V. A. Oboznov, A. Yu. Rusanov, A. V. Veretennikov, A. A. Golubov, and J. Aarts, Coupling of two superconductors through a ferromagnet: evidence for a $\pi$ junction, Phys. Rev. Lett. \textbf{86}, 2427 (2001).
\bibitem{Kontos}
T. Kontos, M. Aprili, J. Lesueur, F. Gen$\hat{\text{e}}$t, B. Stephanidis, and R. Boursier, Josephson Junction through a Thin Ferromagnetic Layer: Negative Coupling, Phys. Rev. Lett. \textbf{89}, 137007 (2002).
\bibitem{Gingrich}
E. C. Gingrich, B. M. Niedzielski, J. A. Glick, Y. Wang, D. L. Miller, R. Loloee, W. P. Pratt Jr, and N. O. Birge, Controllable $0$-$\pi$ Josephson junctions containing aferromagnetic spin valve, Nat. Phys. \textbf{12}, 564 (2016).
\bibitem{Yuan}
N. F. Q. Yuan and L. Fu, Supercurrent diode effect and finite-momentum superconductors, Proc. Natl. Acad. Sci. USA \textbf{119}, e2119548119 (2022).
\bibitem{Ando}
F. Ando, Y. Miyasaka, T. Li, J. Ishizuka, T. Arakawa, Y. Shiota, T. Moriyama, Y. Yanase, and T. Ono, Observation of superconducting diode effect, Nature (London) \textbf{584}, 373 (2020).
\bibitem{Nitta}
J. Nitta, T. Akazaki, H. Takayanagi, and T. Enoki, Gate control of spin-Orbit interaction in an inverted In$_0.53$Ga$_0.47$As/In$_0.52$Al$_0.48$As heterostructure, Phys. Rev. Lett. \textbf{78}, 1335 (1997).
\bibitem{Shalom}
M. Ben Shalom, M. Sachs, D. Rakhmilevitch, A. Palevski, and Y. Dagan, Tuning spin-Orbit coupling and superconductivity at the SrTiO$_3$/LaAlO$_3$ interface: a magnetotransport study, Phys. Rev. Lett. \textbf{104}, 126802 (2010).
\end{thebibliography}
\end{document}